\documentclass[sigconf, nonacm=true,authorversion=true]{acmart}
\usepackage{graphicx} %
\usepackage{xcolor}

\usepackage{ifthen}
\newboolean{showcomments}
\setboolean{showcomments}{true}
\ifthenelse{\boolean{showcomments}}
{ \newcommand{\mynote}[3]{
		\fbox{\bfseries\sffamily\scriptsize#1}
		{\small$\blacktriangleright$\textsf{\emph{\color{#3}{#2}}}$\blacktriangleleft$}}
	\newcommand{\zzz}[1]{{\setlength{\fboxsep}{2pt}\fcolorbox{black}{yellow}{\textsf{\emph{#1}}}}\xspace}}
{ \newcommand{\mynote}[3]{}
	\newcommand{\zzz}[1]{}}

\definecolor{maroon}{rgb}{0.5, 0.0, 0.0}

\usepackage{acronym}
\acrodef{ML}{machine learning}
\acrodef{RCA}{root cause analysis}
\acrodef{IaC}{infrastructure as code}
\acrodef{LLM}{large language model}
\usepackage{graphicx}
\graphicspath{ {figures/} }

\usepackage{algorithm}
\usepackage{algpseudocode}
\usepackage[algo2e,ruled,lined,boxed,commentsnumbered, noend, linesnumbered, procnumbered]{algorithm2e}
\usepackage{cleveref}
\usepackage{siunitx}
\usepackage{booktabs}
\usepackage{multirow}
\usepackage{listings}
\usepackage{subcaption}

\usepackage{tikz}
\usepackage{diagbox}
\usepackage{makecell}
\usepackage[inline]{enumitem}

\usepackage[abbreviations]{foreign}
\usepackage[table]{xcolor}
\usepackage{colortbl}
\usepackage{float}
\restylefloat{table}

\usepackage{tikz}
\usepackage{pgfplots}
\usepackage{pgfplotstable}
\usepackage[eulergreek]{sansmath}
\usepackage{comment}
\newcolumntype{L}[1]{>{\raggedright\let\newline\\\arraybackslash\hspace{0pt}}m{#1}}
\newcolumntype{C}[1]{>{\centering\let\newline\\\arraybackslash\hspace{0pt}}m{#1}}
\newcolumntype{R}[1]{>{\raggedleft\let\newline\\\arraybackslash\hspace{0pt}}m{#1}}

\pgfplotsset{compat=newest}
\usepgfplotslibrary{external,units,colorbrewer,groupplots,fillbetween}
\tikzsetexternalprefix{figures/}
\tikzset{external/mode=list and make}
\usetikzlibrary{patterns}
\usetikzlibrary{shapes.geometric, arrows, positioning}

\newcommand{\newgroupwidth}[2]%
{\expandafter\xdef\csname groupwidth#1\endcsname{#2}}

\newcounter{groupwidth}
\newsavebox{\groupwidthbox}
\makeatletter
{\edef\groupnumber{#1}%
	\stepcounter{groupwidth}%
	\@ifundefined{groupwidth\thegroupwidth}{\pgfmathsetlengthmacro{\mywidth}{\linewidth/\groupnumber}}%
	{\expandafter\let\expandafter\mywidth\csname groupwidth\thegroupwidth\endcsname}%
	\begin{lrbox}{\groupwidthbox}%
		\tikzset{/pgfplots/width={\mywidth}}%
		\ignorespaces}%
	{\end{lrbox}%
	\usebox\groupwidthbox
	\pgfmathsetlengthmacro{\mywidth}{\mywidth + (\linewidth - \wd\groupwidthbox)/\groupnumber}
	\immediate\write\@auxout{\string\newgroupwidth{\thegroupwidth}{\mywidth}}}
\makeatother
\usepackage{physics}
\usepackage{thm-restate}
\usepackage{bbm}

\allowdisplaybreaks

\newcommand{\sys}{\textsc{RIVA}\xspace}
\newcommand{\aiopslab}{\textsc{AIOpsLab}\xspace}
\usepackage{booktabs}
\usepackage{multirow}
\usepackage[table]{xcolor}
\usepackage{graphicx}

\title{\sys: Leveraging LLM Agents for Reliable Configuration Drift Detection}
\date{}

\author{Sami Abuzakuk}
\orcid{0009-0003-6207-5905}
\affiliation{
  \institution{EPFL}
  \city{Lausanne}
  \country{Switzerland}
}
\email{sami.abuzakuk@epfl.ch}

\author{Lucas Crijns}
\affiliation{
  \institution{CYD Campus\\armasuisse W+T}
  \city{Lausanne}
  \country{Switzerland}
}
\email{lucas.crijns@ar.admin.ch}

\author{Anne-Marie Kermarrec}
\orcid{0000-0001-8187-724X}
\affiliation{
  \institution{EPFL}
  \city{Lausanne}
  \country{Switzerland}
}
\email{anne-marie.kermarrec@epfl.ch}

\author{Rafael Pires}
\orcid{0000-0002-7826-1599}
\affiliation{
  \institution{EPFL}
  \city{Lausanne}
  \country{Switzerland}
}
\email{rafael.pires@epfl.ch}

\author{Martijn de Vos}
\orcid{0000-0003-4157-4847}
\affiliation{
  \institution{EPFL}
  \city{Lausanne}
  \country{Switzerland}
}
\email{martijn.devos@epfl.ch}

\makeatletter
\AtBeginDocument{%
	\def\ltx@label#1{\cref@label{#1}}%

	\def\label@in@display@noarg#1{\cref@old@label@in@display{#1}}%

	\def\label@in@mmeasure@noarg#1{%
		\begingroup
		\measuring@false
		\cref@old@label@in@display{#1}%
		\endgroup
	}%
}
\makeatother

\begin{abstract}
\Ac{IaC} tools automate cloud provisioning but verifying that deployed systems remain consistent with the \ac{IaC} specifications remains challenging.
Such \emph{configuration drift} occurs because of bugs in the \ac{IaC} specification, manual changes, or system updates.
\Ac{LLM}-based agentic AI systems can automate the analysis of large volumes of telemetry data, making them suitable for the detection of configuration drift.
However, existing agentic systems implicitly assume that the tools they invoke always return correct outputs, making them vulnerable to erroneous tool responses.
Since agents cannot distinguish whether an anomalous tool output reflects a real infrastructure problem or a broken tool, such errors may cause missed drift or false alarms, reducing reliability precisely when it is most needed.
We introduce \sys (Robust Infrastructure by Verification Agents), a novel multi-agent system that performs robust \ac{IaC} verification even when tools produce incorrect or misleading outputs.
\sys employs two specialized agents, a verifier agent and a tool generation agent, that collaborate through iterative cross-validation, multi-perspective verification, and tool call history tracking.
Evaluation on the \textsc{AIOpsLab} benchmark demonstrates that \sys, in the presence of erroneous tool responses, recovers task accuracy from 27.3\% when using a baseline ReAct agent to 50.0\% on average.
\sys also improves task accuracy 28\% to 43.8\% without erroneous tool responses.
Our results show that cross-validation of diverse tool calls enables more reliable autonomous infrastructure verification in production cloud environments.
 \end{abstract}

\begin{document}

\maketitle

\section{Introduction}

\ac{IaC} tools have become ubiquitous in modern IT operations, enabling automated and scalable infrastructure management~\cite{morris2020infrastructure}.
Popular \ac{IaC} tools such as \textsc{Terraform}~\cite{terraform}, \textsc{Ansible}~\cite{ansible}, and \textsc{AWS CloudFormation}~\cite{cloudformation} enable organizations to define, provision, and maintain their infrastructure through version-controlled configuration files, significantly reducing manual effort and configuration errors~\cite{rahman2019systematic,kumara2021s,opdebeeck2025analysing}.
Instead of manually configuring cloud resources, \ac{IaC} engineers describe the desired system state in declarative templates or scripts.
These definitions are then processed by the IaC tool, which compares the declared state with the current environment and automatically applies the necessary changes to reach the target state.
This approach ensures consistency across different environments (\eg, development or production) and makes infrastructure reproducible, testable, and traceable~\cite{verdet2023exploring}.
The \ac{IaC} paradigm is used by major companies such as Amazon, Netflix, Google and Facebook~\cite{morris2020infrastructure}.

\begin{figure}[b]
    \centering
    \includegraphics{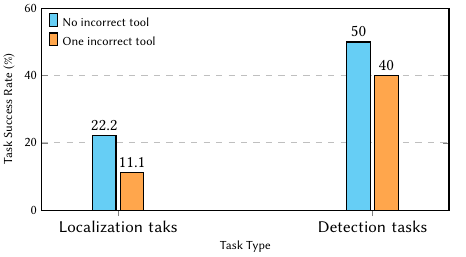}
    \caption{The impact of erroneous tools on the accuracy of error detection and localization tasks in the \textsc{AIOpsLab} benchmark by agentic AI systems.}
    \label{fig:motivation}
\end{figure}

Despite these advantages, verifying that deployed infrastructure adheres to its specification remains a critical challenge~\cite{wang2022infrastructure}.
The main problem is that during or after provisioning, the infrastructure may diverge from the intended state, \ie, configuration drift~\cite{pahl2025infrastructure}.
During provisioning, subtle bugs in \ac{IaC} modules, provider plugins, or cloud APIs can leave resources in a partially created or an inconsistent state, such as virtual machines being provisioned without the correct security groups or permissions only being applied to a subset of expected services~\cite{drosos2024your}.
Such bugs can propagate at scale, potentially leading to wide-scale service disruptions or vulnerabilities over time~\cite{rahman2018bugs,thiyagarajan2024ai}.
Post-deployment, configuration drift introduces further complexities: automated software updates, manual changes by engineers that bypass \ac{IaC} (\eg, in response to an emergency), or performance tuning can modify resource properties in ways that violate the \ac{IaC} specifications~\cite{yang2025automated}.
Detecting configuration drift requires continuous monitoring and correlating massive volumes of logs, events, and state changes to determine whether the live environment still matches the intention of the IaC definitions.
This verification effort is extremely labor-intensive and error-prone~\cite{thiyagarajan2024ai}.

Recent advances in \ac{LLM}-based agentic systems have shown strong potential to improve cloud reliability~\cite{shetty2024building,yang2025cloud}.
In the context of \ac{IaC} and configuration drift, agentic AI systems can interpret logs, events, and configuration states, summarize large volumes of  telemetry data, and autonomously decide when and how to investigate anomalies~\cite{thiyagarajan2024ai}.
By leveraging \acp{LLM}’s abilities in pattern recognition, reasoning, and cross-correlating information from different sources, these agents can detect configuration drifts that are difficult to detect with traditional rule-based tools.
Moreover, agentic workflows can iteratively refine hypotheses, query cloud APIs, and validate suspected issues against the \ac{IaC} specification.
This shows high potential to reduces manual effort, shorten incident-detection times, and to provide operators with actionable, high-level summaries of the configuration drift.

A key vulnerability of agentic AI in the context of \ac{IaC}, however, is their implicit assumption that the tools they invoke always return correct and trustworthy outputs.
In practice, IaC-related tools may fail or behave inconsistently due to API outages, throttling, transient network issues, parsing errors, or outdated provider implementations.
Crucially, such erroneous tool responses are sometimes indistinguishable from genuine infrastructure misconfigurations: both manifest as unexpected states, missing fields, or contradictory information.
As a result, an agent that relies solely on tool feedback cannot determine whether it is observing a real configuration drift or merely a faulty tool invocation.
In~\Cref{fig:motivation} we show the impact of a single erroneous tool on the success rate of localization and detection tasks in the \aiopslab benchmark~\cite{chen2025aiopslab}.
The presence of erroneous tools, silently returning incorrect information, negatively impact the task success rate, from 22.2\% to 11.1\% and 50.0\% to 40.0\% for the localization and detection tasks, respectively.
Existing agentic solutions lack mechanisms for detecting or reasoning about these inaccuracies in tool outputs.

This work introduces \sys (Robust Infrastructure by Verification Agents), a novel multi-agent system designed to perform robust IaC-based infrastructure verification even when the tools it relies on produce incorrect or misleading outputs.
The main insight of \sys is that it is unlikely that two tool calls with the same goal will both be erroneous.
Our system enables agents with the ability to detect, reason about, and overcome unreliable tool responses through iterative cross-validation, multi-perspective verification, and history tracking of tool results.
At the core of \sys are two specialized agents, a verifier and tool generation agent.
Together, these agents converges on reliable conclusions about infrastructure compliance.
Moreover, these agents have access to a tool call history which they use to generate and invoke unique tool calls to verify a single property in the \ac{IaC} specifications.

We implement and evaluate our system on the \aiopslab benchmark.
We compare the performance of \sys against that of a ReAct agent, which is a single agent that alternates between reasoning and tool calls.
In the presence of unreliable tools, \sys across all tasks recovers the task success rate from 27.3\% when using a baseline ReAct agent to 50.0\%.
Furthermore, our results show superior accuracy and efficiency across all tasks even when tools generate correct responses, improving the average task success rate from 28\% to 43.8\%.
Overall, we find that \sys is an effective solution to mitigate the effect of incorrect tools and provide robust infrastructure verification.

\textbf{Contributions.}
Our contributions are as follows:
\begin{itemize}
    \item We design \sys, a novel multi-agent system that is able to detect configuration drifts in the presence of erroneous tool call outputs (\Cref{sec:design}).
    \item We implement \sys and integrate our system into \aiopslab, an open-source framework for the structured evaluation of AI agents (\Cref{sec:exp_setup}).
    \item We evaluate the effectiveness of \sys and show that it significantly outperforms a ReAct agent, in terms of task success rate, on tasks in the \aiopslab benchmark, both with and without erroneous tool responses (\Cref{sec:exp}).
    This comes with a manageable overhead in number of tokens used.
\end{itemize}

\section{Background and Problem Description}
\label{sec:background}
We next provide background on \ac{IaC}, configuration drift and agentic AI.
We then formulate the main research challenge that this work addresses.

\subsection{\ac{IaC} and Configuration Drift}
\ac{IaC} tools such as Terraform, Ansible, and AWS CloudFormation allow operators to define the intended state of their infrastructure declaratively, enabling reproducibility, automation, and version control.
In practice, however, deployed resources frequently diverge from the \ac{IaC} definition—a phenomenon known as \emph{configuration drift}~\cite{pahl2025infrastructure}.
Drift can arise during provisioning when subtle bugs or provider API inconsistencies cause resources to be created with missing or incorrect attributes without any explicit error being reported, or after provisioning when manual hotfixes, automated scaling actions, or emergency interventions modify parameters outside the \ac{IaC} workflow.
Over time, these discrepancies accumulate, making the operational state increasingly disconnected from its specification, and detecting them at cloud scale requires extensive tooling and automation.
A concrete \textsc{Terraform} example is provided in Appendix~\ref{app:terraform}.

\subsection{Agentic AI}
\Acp{LLM} have recently emerged as a compelling solution for cloud engineering and improved reliability~\cite{parthasarathy2025engineering,xiang2025automated,shetty2024building}.
In particular, \acp{LLM} are able to process the heterogeneous and unstructured telemetry data that are common in modern cloud environments~\cite{roy:2024:llm:rca}.
Their strengths in pattern recognition, summarization, and contextual reasoning make them well suited for interpreting complex system behavior and identifying potential issues, \eg, the detection of concept drift.
However, using an \ac{LLM} in isolation is not sufficient: effective infrastructure verification requires iterative reasoning and continuous interaction with the live environment to fetch the appropriate data from nodes and services.

To enable such interaction, there is a shift towards adopting \emph{agentic AI} where autonomous agents coordinate with external tools such as cloud APIs, telemetry endpoints, or custom scripts, to gather evidence and refine their understanding of the system state~\cite{hughes2025ai,murugesan2025rise}.
More specifically, tools are external service interfaces that agents can programmatically invoke to execute functions beyond the inherent capabilities of the underlying \acp{LLM}, thus extending the capabilities of \Acp{LLM} by providing direct programmatic access to telemetry and configuration data.
A central part of the \ac{LLM} agent’s decision-making process is the selection, invocation, and orchestration of the available tools~\cite{schick2023toolformer}.

In the context of \ac{IaC} and configuration drift, tool calls allow agents to probe the environment, query the properties of deployed resources, or retrieve telemetry needed to confirm whether the actual system matches the IaC specification.
A detailed verification workflow is illustrated in Appendix~\ref{app:workflow}.
A widely used approach for structuring such interactions is the \textsc{ReAct} framework, in which an agent alternates between natural-language reasoning steps and concrete tool invocations~\cite{yao2023react}.
We refer to the resulting sequence of reasoning and tool-use decisions as a \emph{trajectory}. %
In summary, agentic AI provides a promising direction for addressing the challenges posed by configuration drift.

\subsection{Problem Description}
The effectiveness of agentic AI in the context of \ac{IaC} verification fundamentally depends on the reliability of the tools they invoke.
When tools provide incomplete, stale, or misleading outputs, agents may incorrectly conclude that a resource is misconfigured or, conversely, that the infrastructure is healthy even when configuration drift has occurred.
This motivates the need for \emph{robust agentic frameworks} that are capable of successful operation even when the correctness of tool outputs cannot be guaranteed.

An illustrative example of how erroneous tool outputs can silently mislead an agent can be found in Appendix~\ref{app:pingnode}.

In summary, the core problem is that agentic AI systems rely on tool outputs that may themselves be unreliable, incomplete, or stale, and current architectures provide no principled mechanism to detect or reason about such inconsistencies.
As a result, agents cannot reliably distinguish genuine configuration drift from faulty tool behavior.
This leads to the central research question of this work: \emph{How can we design agentic AI systems that robustly verify IaC-defined infrastructure even when the tools they depend on return incorrect or misleading outputs?}

\section{Design of \sys}
\label{sec:design}

\begin{figure*}[ht]
\centering
\includegraphics[width=.85\linewidth]{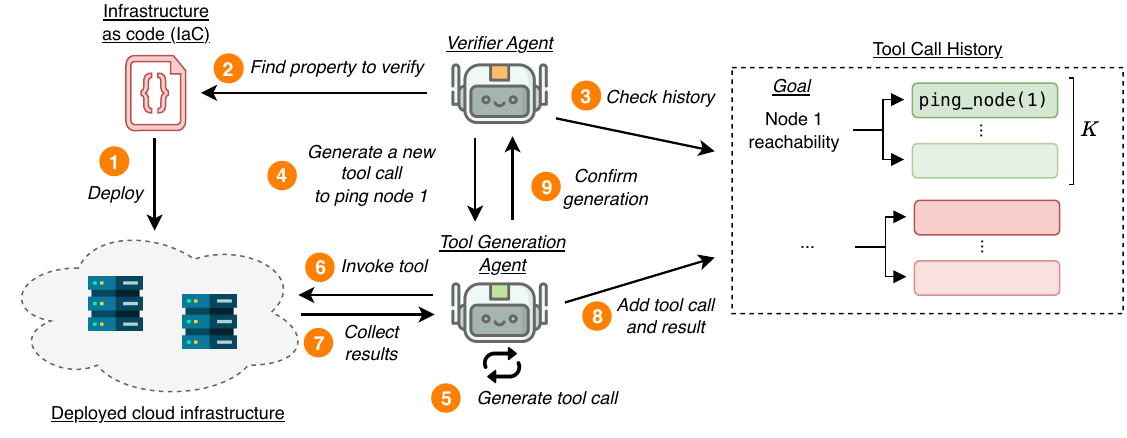}
\caption{The \sys system architecture and workflow.}
\label{fig:system_design}
\end{figure*}

We design \sys, a multi-agent architecture for \ac{IaC} verification in the presence of unreliable tools.
We visualize the \sys system architecture and workflow in~\Cref{fig:system_design}.
At the core of our system architecture are two collaborating \ac{LLM} agents: a \emph{Verifier Agent} and a \emph{Tool Generation Agent}.
The verifier agent has access to the \ac{IaC} specification and finds properties to verify.
The tool generation agent generates tool calls, executes these tool calls, collects the results, and sends the result back to the verifier agent.

The main insight behind \sys is that even though a single tool call may silently return incorrect or misleading information, it is highly unlikely that multiple independent tools will all fail in the same way.
In particular, tools are usually implemented as wrappers around complex commands in order to simplify the interaction with the agent.
For example, in a Kubernetes environment, one might implement a \texttt{get\_logs} tool using the Kubernetes executable, abstracting away the complexity of using this binary.
However, accessing the logs of a pod can also be done by directly accessing the log files in the pod (i.e., with SSH).
A subtle drift in the infrastructure configuration might render the original \texttt{get\_logs} invalid but would probably not impact a direct access to the machine.

Thus, instead of trusting individual tool responses, we verify each property through cross-validation across diverse tool calls.
By comparing their outputs, identifying contradictions, and iteratively refining verification attempts, an agent can distinguish genuine configuration drift from erroneous tool behavior.
Therefore, both the verifier agent and tool generation agent have access to a shared data structure called the Tool Call History to verify infrastructure compliance with specifications.

In the following, we first elaborate on each component in our system and then describe the full \sys workflow.

\subsection{Tool History}

The Tool History serves as the central data structure enabling agent collaboration and tool reliability assessment.
It is implemented as a map where each key corresponds to a property identifier (a specific line in the infrastructure specification), and each value is an array containing up to $K$ tool execution records.
Every execution record is required to use a different tool.
For example, with $K=2$ and testing the reachability of node 1, the tool generation agent could add an execution record for \texttt{ping\_node(id=1)} and \texttt{send\_message(id=1)}.
Effectively, the hyperparameter $K$ defines the number of diagnostic paths that must be executed for the Verifier Agent to reach a conclusion.
We evaluate the impact of this parameter on the task success rate in \Cref{ssec:k}.

Each record in the tool history contains three elements: (1) the executed command with its arguments, (2) the results generated by the tool, and (3) a brief analysis explaining what the results indicate about the property.
This structure allows the Verifier Agent to examine multiple verification attempts for the same property, facilitating the detection of inconsistent or unreliable tool outputs through cross-validation.

\subsection{Tool Generation Agent}

The Tool Generation Agent generates verification commands tailored to specific properties.
When invoked, the agent first examines the Tool History to identify all previously executed tools for the target property.
It then synthesizes a new verification approach that employs a different technique or checks the property from an alternative angle, ensuring diversity in verification methods.
After generating the command, the agent executes it on the deployed infrastructure and collects the results.
If the tool call is incorrect (wrong arguments, syntax errors, etc.), the agent corrects the issue and executes the command again.
Finally, when the tool successfully runs on the infrastructure, the agent creates a new entry in the Tool History containing the command specification, the execution results, and an analysis that interprets what these results reveal about the property's compliance status.

\subsection{Verifier Agent}

The Verifier Agent is responsible for validating that the deployed infrastructure adheres to the system specification.
For each property in the specification, the agent queries the Tool History to determine whether verification attempts exist for that property.
If no entry is found, the agent requests the Tool Generation Agent to create an appropriate verification command.
If entries exist, the agent analyzes the stored tool results to determine whether they provide sufficient evidence to conclude that the property is satisfied or violated.
Crucially, the agent formulates a conclusion about the property only if $K$ entries are available (i.e., if $K$ distinct diagnostic paths were explored to generate results).
While preventing the agent from reaching a conclusion too quickly, this design still allows the agent to validate other properties.
In particular, because the tool generation agent generates one tool call at a time, the verifier agent has a reasoning step interleaved between each generation.
This allows the agent to reevaluate its plan frequently and, for example, in situations where the initially selected property is not strictly necessary to move forward in the verification task, add a new goal to the tool history and abandon the first goal.
Note that in such a case, because of the $K$ constraint, the abandoned goal will not be considered as proven or disproven by the agent.
\section{Experimental Setup}
\label{sec:exp_setup}

We now describe the experimental setup used to evaluate \sys, including the \aiopslab benchmark, the implementation details of \sys, and the evaluation process.

\subsection{The \aiopslab Benchmark}

We conduct our evaluation using \aiopslab, an open-source framework developed by Microsoft Research for designing, developing, and evaluating autonomous AIOps agents~\cite{chen2025aiopslab}.
\aiopslab provides a holistic infrastructure that can deploy microservice cloud environments, inject faults, generate workloads, and export comprehensive telemetry data.
Moreover, \aiopslab supports several critical AIOps tasks including incident detection, localization, root cause diagnosis, and mitigation, making it an ideal testbed for evaluating infrastructure verification by \ac{LLM} agents.

\subsection{Baseline Agent}

For comparison, we use the ReAct agent implementation in \aiopslab as our baseline.
The ReAct framework enables agents to interleave reasoning and action steps, allowing them to generate verbal reasoning traces before selecting and executing tools~\cite{yao2023react}.
We select ReAct as the baseline because it heavily relies on tool invocations and is the standard pattern adopted by most existing agentic AI systems~\cite{schick2023toolformer}.

\subsection{Tool Reliability Evaluation}

To evaluate the effectiveness of our approach in handling incorrect tool outputs, we modify the behavior of some observability tools provided by \aiopslab.
The \aiopslab implementation exposes various tools to collect telemetry data from deployed infrastructure.
Specifically, we configure the \texttt{get\_logs}, \texttt{read\_traces}, and \texttt{read\_metrics} functions to return empty strings rather than the expected telemetry data.
These modifications simulate tool unreliability while maintaining the distinction between incorrect outputs and genuine errors, tool invocations that violate the expected interface or encounter execution failures still raise exceptions as expected.
This design choice reflects scenarios where tools execute successfully but return misleading or incomplete information due to configuration issues, API timeouts, or data collection failures.
To ensure fair comparison between both implementations, the maximum number of steps is set to 45 for each.
For \sys, a multi-agent system, this limit represents the combined steps executed by both agents.

\subsection{Evaluation Protocol}

We evaluate both the baseline ReAct agent and our proposed system across all root cause analysis tasks available in \aiopslab.
The benchmark contains three types of tasks: \emph{localization tasks} identify where the root cause of a particular fault lies, \emph{detection tasks} aim to determine whether an incident has occurred, and \emph{analysis tasks}  diagnose the root cause of identified incidents.
For each task category, we conduct experiments under different tool reliability conditions, systematically introducing incorrect behavior in one the two observability tools (\texttt{get\_logs}, \texttt{read\_metrics}).
We run each task 5 times and measure the accuracy of both systems in correctly completing each task type.
For all experiments, we use the model \texttt{gpt-oss:120b}~\cite{agarwal2025gpt}.
\section{Experimental Evaluation}
\label{sec:exp}

We evaluate the performance of \sys.
Our evaluation answers the following questions:
\begin{itemize}
    \item What is the task success rate of \sys and the ReAct baseline agent on tasks in the \aiopslab benchmark, both with and without erroneous tool responses (\Cref{ssec:accuracy})?
    \item What is the step count and number of tokens used for \sys and the ReAct baseline agent on tasks in the \aiopslab benchmark (\Cref{ssec:efficiency})?
    \item How does the tool call history size (parameterized by $K$) impact the performance of \sys (\Cref{ssec:k})?
\end{itemize}

\subsection{Task Success Rate of \sys and ReAct agents}
\label{ssec:accuracy}

\begin{figure*}[th]
    \centering
    \begin{subfigure}[b]{0.45\textwidth}
        \centering
        \includegraphics{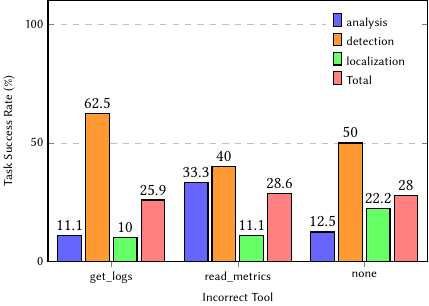}
        \caption{The task success rate of a ReAct agent.}
        \label{fig:total_acc_comparison}
    \end{subfigure}
    \begin{subfigure}[b]{0.45\textwidth}
        \centering
        \includegraphics{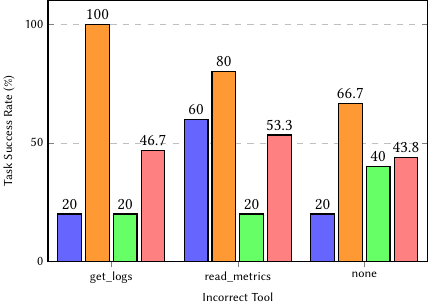}
        \caption{The task success rate of \sys.}
        \label{fig:riva_breakdown}
    \end{subfigure}
    \caption{The task success rate of \sys (with $K=2$) and a ReAct agent in the presence of erroneous tools and without any erroneous tool.}
    \label{fig:comparison_figure}
\end{figure*}

We run \sys (with $K=2$) and the ReAct agent on all detection, localization, and analysis tasks in the \aiopslab benchmark.
\Cref{fig:comparison_figure} shows the task success rate of each task type with and without erroneous tools.
In the presence of incorrect tools, the task success rate of \sys is always greater than or equal to that of the ReAct agent with only correct tools.
For instance, the accuracy of localization tasks with the ReAct implementation is 22.2\% with correct tools and 11.1\% with an incorrect \texttt{get\_logs} tool implementation.
\sys achieves an accuracy of 20.0\% with the same incorrect tool.
Notably, across all tasks, \sys improves the average accuracy from 28.0\% to 43.8\% even when all tool responses are correct.
Careful analysis of the agentic trajectories reveals that most of this improvement stems from the multi-agent design of \sys.
By separating responsibilities, the verifier agent focuses exclusively on interpreting tool call results rather than managing their creation and correction.
This division of labor reduces reasoning errors that plague ReAct agents, which often lose sight of their initial goal after saturating their context with failed tool executions.

These results show that \sys can reach superior task success rates even with erroneous tool responses.

However, \sys fails to reach accuracies similar to those achieved when executed with correct tools.
This suggests that our system does not fully mitigate the impact of erroneous tools.
In particular, for localization tasks, \sys reaches 40.0\% task success rate when all tool calls are correct but achieves 20.0\% when the \texttt{get\_logs} and \texttt{read\_metrics} tools are incorrect.
An inspection of several trajectories reveals that the tool generation agent is responsible for this loss in accuracy.
Specifically, because this agent generates tool calls, it is responsible for correcting incorrect tool calls.
For example, if the agent fails to identify a correct service name, it will attempt to find the correct service name by itself using its tools.
These tool calls are never added to the tool history, meaning that subsequent calls to the tool generator will likely have to redo the same work.
This increases the number of steps needed for each generation and the probability of error, leading to decreased accuracy.
This issue could be mitigated by dynamically appending exploratory tool calls to the history, enabling the tool generation agent to build upon previous correction attempts rather than repeating the same discovery process.

Notably, the presence of erroneous tool calls \emph{sometimes improves the accuracy of the ReAct agent}.
For example, the task success rate of the ReAct agent for detection tasks is 50.0\% whereas the success rate with an incorrect \texttt{get\_logs} reaches 62.5\%.
We note that detection tasks are the easiest tasks as they can be solved by only finding an irregular metric or log.
Thus, by making \texttt{get\_logs} return an empty string, the base agent ignores the results and immediately focuses on the metric results.
Metrics being more focused and less verbose lead to fewer hallucinations and thus better results.
Overall, the results in~\Cref{fig:comparison_figure} suggest that \sys successfully leverages its design to mitigate the effect of incorrect tools.

\subsection{Efficiency of \sys and ReAct agents}
\label{ssec:efficiency}

We compare the efficiency of \sys and ReAct by quantifying trajectory length and token usage, which directly capture the duration and computational cost of agentic reasoning.
With correct tools, 80\% of \sys ($K=2$) tasks complete within 15 steps versus 60\% for ReAct; \sys's step count peaks at 17 while 33\% of ReAct runs hit the 45-step cap, and peak token usage is \num{38000} for \sys versus \num{78000} for ReAct.
We attribute this to the two-agent design of \sys, which distributes context across agents, reducing hallucination risk and the number of corrective steps.
With erroneous tools, \sys still requires at most 17 steps whereas 37\% of ReAct tasks exceed this threshold; peak token usage is \num{50000} for \sys versus \num{90000} for ReAct.
Detailed CDF distributions are shown in Appendix~\ref{app:efficiency}.

\subsection{The Impact of $K$ on Efficiency}
\label{ssec:k}

\begin{table}
\centering
\begin{tabular}{|l|c|c|c|}

\hline
Agent & Only correct & incorrect & incorrect \\
& tools & get\_logs & read\_metrics \\
\hline
ReAct       & 28.00 & 25.93 & 28.57 \\
\sys (K=1)  & 27.67 & 24.80 & 29.02 \\
\sys (K=2)  & 43.75 & 46.67 & 53.33 \\
\sys (K=3)  & 0     & 0     & 0 \\
\hline
\end{tabular}
\caption{Average success rate of ReAct and \sys with different $K$ on \aiopslab}
\label{tab:k_analysis}
\end{table}

We now examine the impact of tuning the hyperparameter $K$, which governs the number of unique tool calls necessary to reach a conclusion.
The average success rate of ReAct and \sys with $K$ set to 1, 2 and 3 are shown in \Cref{tab:k_analysis}.

When $K$ is set to 1, \sys achieves similar numbers to ReAct.
By using only 1 tool call to validate a property, the advantages of \sys are effectively negated and it falls into the same trap as ReAct.
When $K$ is set to 2, as seen in previous subsections, \sys outperforms ReAct and is able to avoid the accuracy degradation caused by the incorrect tool call.
However, when $K$ is higher than 2, the accuracy falls to 0\%.
This failure stems from a fundamental constraint of AIOpsLab: it does not offer enough diagnostic paths for the generator agent to produce three distinct tool calls for the same goal, causing it to loop until hitting the step limit.
A detailed analysis of this failure mode is provided in Appendix~\ref{app:k}.

These results highlight that \sys performance depends critically on hyperparameter $K$.
If the system cannot generate at least $K$ distinct tool calls to verify the same property, it will fail to reach a conclusion.
This hyperparameter directly reflects the diagnostic flexibility afforded to the agent.
In constrained environments like \aiopslab, where the agent has limited alternative diagnostic paths, $K$ should be set lower accordingly.
This analysis also reveals a promising direction for improving \sys design: when the Tool Generation agent cannot identify $K$ distinct paths, the system could continue its analysis with reduced confidence levels rather than halting entirely.
\section{Related Work}

The challenge of ensuring robust agent execution in the presence of tool errors has received growing attention in recent research on \ac{LLM}-based agentic systems.
Approaches generally fall into two categories: handling explicit execution failures and mitigating subtle, silent errors.

Vuddanti et al. introduce PALADIN~\cite{vuddanti2025paladin}, a self-correcting language model agent designed to handle tool malfunctions such as timeouts and API exceptions.
Their approach focuses on enabling agents to recover from tool execution failures through self-correction mechanisms.
However, PALADIN assumes that tool failures manifest as explicit exceptions (execution errors).

Similarly, Sheffler proposes a temporal expression language for monitoring agent behavior and detecting deviations from expected behavioral patterns~\cite{sheffler2025approach}.
This approach monitors execution traces of tool calls and state transitions to identify anomalies in agent actions, such as improper tool sequencing and failed coordination.
While effective for detecting behavioral regressions, this method requires predefined temporal patterns and does not explicitly address the problem of verifying tool output correctness.

As opposed to explicit failures, the challenge of silent errors, where a tool successfully executes but returns an incorrect output, has been highlighted as a critical issue in recent work~\cite{sun-etal-2024-tools}.
Tools Fail investigates methods for LLMs to learn to doubt tools and detect these mistakes, often using in-context interventions like checklists.
While this work confirms the existence and importance of the silent error problem, it primarily focuses on the LLM's capacity for detection based on internal expectations, whereas our work provides a structured, external mechanism for tool output verification.

Some work tackle the tool error problems by improving recovery.
Several approaches explore learning-based error recovery mechanisms, often involving explicit reflection processes that diagnose failed tool calls and propose corrected alternatives~\cite{su2025failure}.
Additionally, AgentDebug provides a debugging framework that isolates root-cause failures and generates corrective feedback, enabling agents to iteratively recover from errors~\cite{zhu2025llm}.

On the side of provable safety, VeriGuard tackles a related challenge by integrating formal verification into the code generation process to ensure 'correct-by-construction' agent policies, using an iterative refinement loop guided by a verifier's counterexamples~\cite{miculicich2025veriguard}.

While these works address various aspects of tool reliability and error handling—from explicit exceptions (PALADIN) to internal detection of silent errors (Tools Fail) and policy-level verification (VeriGuard), they primarily focus on detecting and recovering from explicit tool failures or execution errors.
Our approach differs by specifically targeting scenarios where tools execute without raising exceptions but return incorrect results, requiring the agent to reason about tool output reliability through cross-validation and iterative verification rather than relying on explicit error signals.
\section{Conclusions and Future Work}

This paper presents \sys, a multi-agent system that detects configuration drift in cloud infrastructure even when tools return incorrect data.
By cross-validating multiple independent tool calls targeting the same property, \sys distinguishes real infrastructure problems from faulty tool outputs.
Evaluation on the \textsc{AIOpsLab} benchmark demonstrates that \sys, in the presence of erroneous tool responses, recovers task accuracy from 27.3\% when using a baseline ReAct agent to 50.0\% on average.
\sys also improves task accuracy 28\% to 43.8\% without erroneous tool responses.
Moreover, our system completes most tasks more efficiently, using fewer steps and tokens compared to a baseline ReAct agent.
Thus, \sys demonstrates that cross-validation enables more reliable autonomous infrastructure verification in production environments.

Future work should focus on three key areas to enhance the robustness and applicability of \sys.
First, the tool generation behavior needs improvement to enable the agent to dynamically add more than one tool call to the history when it uses multiple tools to generate the requested tool call.
Second, the overall system reaction to hyperparameter $K$ requires refinement so that the system can continue its analysis even when it fails to find $K$ alternative paths, rather than failing completely.
Third, the \aiopslab benchmark proves too restrictive for a comprehensive analysis of \sys's capabilities, necessitating the development or adoption of a better benchmark that more closely mirrors actual \ac{IaC} deployment scenarios.
This new benchmark should provide direct access to the deployment environment, allowing the agent to explore a much wider range of diagnostic paths and more thoroughly demonstrate \sys's verification capabilities in realistic production settings.

\begin{acks}
This work was partially supported by Cyber Defence Campus (project number AR-CYD-C-024, contract number 8010075905) and the Swiss National Science Foundation, under the project FRIDAY: Frugal, Privacy-Aware and Practical Decentralized Learning, SNSF proposal No. 10.001.796.
We thank Gérôme Bovet, Sayan Biswas, and Milos Vujasinovic for their inputs and insightful discussions.
\end{acks}

\bibliographystyle{ACM-Reference-Format}
\bibliography{references.bib}


\begin{thebibliography}{31}


\ifx \showCODEN    \undefined \def \showCODEN     #1{\unskip}     \fi
\ifx \showISBNx    \undefined \def \showISBNx     #1{\unskip}     \fi
\ifx \showISBNxiii \undefined \def \showISBNxiii  #1{\unskip}     \fi
\ifx \showISSN     \undefined \def \showISSN      #1{\unskip}     \fi
\ifx \showLCCN     \undefined \def \showLCCN      #1{\unskip}     \fi
\ifx \shownote     \undefined \def \shownote      #1{#1}          \fi
\ifx \showarticletitle \undefined \def \showarticletitle #1{#1}   \fi
\ifx \showURL      \undefined \def \showURL       {\relax}        \fi
\providecommand\bibfield[2]{#2}
\providecommand\bibinfo[2]{#2}
\providecommand\natexlab[1]{#1}
\providecommand\showeprint[2][]{arXiv:#2}

\bibitem[Agarwal et~al\mbox{.}(2025)]%
        {agarwal2025gpt}
\bibfield{author}{\bibinfo{person}{Sandhini Agarwal}, \bibinfo{person}{Lama
  Ahmad}, \bibinfo{person}{Jason Ai}, \bibinfo{person}{Sam Altman},
  \bibinfo{person}{Andy Applebaum}, \bibinfo{person}{Edwin Arbus},
  \bibinfo{person}{Rahul~K Arora}, \bibinfo{person}{Yu Bai},
  \bibinfo{person}{Bowen Baker}, \bibinfo{person}{Haiming Bao},
  {et~al\mbox{.}}} \bibinfo{year}{2025}\natexlab{}.
\newblock \showarticletitle{gpt-oss-120b \& gpt-oss-20b model card}.
\newblock \bibinfo{journal}{\emph{arXiv preprint arXiv:2508.10925}}
  (\bibinfo{year}{2025}).
\newblock


\bibitem[{Amazon Web Services}(2025)]%
        {cloudformation}
\bibfield{author}{\bibinfo{person}{{Amazon Web Services}}.}
  \bibinfo{year}{2025}\natexlab{}.
\newblock \bibinfo{title}{AWS CloudFormation}.
\newblock \bibinfo{howpublished}{\url{https://aws.amazon.com/cloudformation/}}.
\newblock
\newblock
\shownote{Accessed: 2025-02-24}.


\bibitem[Chen et~al\mbox{.}(2025)]%
        {chen2025aiopslab}
\bibfield{author}{\bibinfo{person}{Yinfang Chen}, \bibinfo{person}{Manish
  Shetty}, \bibinfo{person}{Gagan Somashekar}, \bibinfo{person}{Minghua Ma},
  \bibinfo{person}{Yogesh Simmhan}, \bibinfo{person}{Jonathan Mace},
  \bibinfo{person}{Chetan Bansal}, \bibinfo{person}{Rujia Wang}, {and}
  \bibinfo{person}{Saravan Rajmohan}.} \bibinfo{year}{2025}\natexlab{}.
\newblock \showarticletitle{Aiopslab: A holistic framework to evaluate ai
  agents for enabling autonomous clouds}.
\newblock \bibinfo{journal}{\emph{arXiv preprint arXiv:2501.06706}}
  (\bibinfo{year}{2025}).
\newblock


\bibitem[Drosos et~al\mbox{.}(2024)]%
        {drosos2024your}
\bibfield{author}{\bibinfo{person}{Georgios-Petros Drosos},
  \bibinfo{person}{Thodoris Sotiropoulos}, \bibinfo{person}{Georgios
  Alexopoulos}, \bibinfo{person}{Dimitris Mitropoulos}, {and}
  \bibinfo{person}{Zhendong Su}.} \bibinfo{year}{2024}\natexlab{}.
\newblock \showarticletitle{When your infrastructure is a buggy program:
  Understanding faults in infrastructure as code ecosystems}.
\newblock \bibinfo{journal}{\emph{Proceedings of the ACM on Programming
  Languages}} \bibinfo{volume}{8}, \bibinfo{number}{OOPSLA2}
  (\bibinfo{year}{2024}), \bibinfo{pages}{2490--2520}.
\newblock


\bibitem[{HashiCorp}(2025)]%
        {terraform}
\bibfield{author}{\bibinfo{person}{{HashiCorp}}.}
  \bibinfo{year}{2025}\natexlab{}.
\newblock \bibinfo{title}{Terraform}.
\newblock \bibinfo{howpublished}{\url{https://www.terraform.io/}}.
\newblock
\newblock
\shownote{Accessed: 2025-02-24}.


\bibitem[Hughes et~al\mbox{.}(2025)]%
        {hughes2025ai}
\bibfield{author}{\bibinfo{person}{Laurie Hughes}, \bibinfo{person}{Yogesh~K
  Dwivedi}, \bibinfo{person}{Tegwen Malik}, \bibinfo{person}{Mazen Shawosh},
  \bibinfo{person}{Mousa~Ahmed Albashrawi}, \bibinfo{person}{Il Jeon},
  \bibinfo{person}{Vincent Dutot}, \bibinfo{person}{Mandanna Appanderanda},
  \bibinfo{person}{Tom Crick}, \bibinfo{person}{Rahul De’}, {et~al\mbox{.}}}
  \bibinfo{year}{2025}\natexlab{}.
\newblock \showarticletitle{AI agents and agentic systems: A multi-expert
  analysis}.
\newblock \bibinfo{journal}{\emph{Journal of Computer Information Systems}}
  (\bibinfo{year}{2025}).
\newblock
\href{https://doi.org/10.1080/08874417.2025.2483832}{doi:\nolinkurl{10.1080/08874417.2025.2483832}}


\bibitem[Kumara et~al\mbox{.}(2021)]%
        {kumara2021s}
\bibfield{author}{\bibinfo{person}{Indika Kumara}, \bibinfo{person}{Mart{\'\i}n
  Garriga}, \bibinfo{person}{Angel~Urbano Romeu}, \bibinfo{person}{Dario
  Di~Nucci}, \bibinfo{person}{Fabio Palomba}, \bibinfo{person}{Damian~Andrew
  Tamburri}, {and} \bibinfo{person}{Willem-Jan van~den Heuvel}.}
  \bibinfo{year}{2021}\natexlab{}.
\newblock \showarticletitle{The do’s and don’ts of infrastructure code: A
  systematic gray literature review}.
\newblock \bibinfo{journal}{\emph{Information and Software Technology}}
  \bibinfo{volume}{137} (\bibinfo{year}{2021}), \bibinfo{pages}{106593}.
\newblock


\bibitem[Miculicich et~al\mbox{.}(2025)]%
        {miculicich2025veriguard}
\bibfield{author}{\bibinfo{person}{Lesly Miculicich}, \bibinfo{person}{Mihir
  Parmar}, \bibinfo{person}{Hamid Palangi}, \bibinfo{person}{Krishnamurthy~Dj
  Dvijotham}, \bibinfo{person}{Mirko Montanari}, \bibinfo{person}{Tomas
  Pfister}, {and} \bibinfo{person}{Long~T Le}.}
  \bibinfo{year}{2025}\natexlab{}.
\newblock \showarticletitle{VeriGuard: Enhancing LLM Agent Safety via Verified
  Code Generation}.
\newblock \bibinfo{journal}{\emph{arXiv preprint arXiv:2510.05156}}
  (\bibinfo{year}{2025}).
\newblock


\bibitem[Morris(2020)]%
        {morris2020infrastructure}
\bibfield{author}{\bibinfo{person}{Kief Morris}.}
  \bibinfo{year}{2020}\natexlab{}.
\newblock \bibinfo{booktitle}{\emph{Infrastructure as code}}.
\newblock \bibinfo{publisher}{O'Reilly Media}.
\newblock


\bibitem[Murugesan(2025)]%
        {murugesan2025rise}
\bibfield{author}{\bibinfo{person}{San Murugesan}.}
  \bibinfo{year}{2025}\natexlab{}.
\newblock \showarticletitle{The Rise of Agentic AI: Implications, Concerns, and
  the Path Forward}.
\newblock \bibinfo{journal}{\emph{IEEE Intelligent Systems}}
  \bibinfo{volume}{40}, \bibinfo{number}{2} (\bibinfo{year}{2025}).
\newblock
\href{https://doi.org/10.1109/MIS.2025.3544940}{doi:\nolinkurl{10.1109/MIS.2025.3544940}}


\bibitem[Opdebeeck et~al\mbox{.}(2025)]%
        {opdebeeck2025analysing}
\bibfield{author}{\bibinfo{person}{Ruben Opdebeeck}, \bibinfo{person}{Bram
  Adams}, {and} \bibinfo{person}{Coen De~Roover}.}
  \bibinfo{year}{2025}\natexlab{}.
\newblock \showarticletitle{Analysing Software Supply Chains of Infrastructure
  as Code: Extraction of Ansible Plugin Dependencies}. In
  \bibinfo{booktitle}{\emph{2025 IEEE International Conference on Software
  Analysis, Evolution and Reengineering (SANER)}}. IEEE,
  \bibinfo{pages}{181--192}.
\newblock


\bibitem[Pahl et~al\mbox{.}(2025)]%
        {pahl2025infrastructure}
\bibfield{author}{\bibinfo{person}{Claus Pahl}, \bibinfo{person}{Niyazi~Gokberk
  Gunduz}, \bibinfo{person}{Ovg{\"u}m~Can Sezen}, \bibinfo{person}{Ali
  Ghamgosar}, {and} \bibinfo{person}{Nabil El~Ioini}.}
  \bibinfo{year}{2025}\natexlab{}.
\newblock \showarticletitle{Infrastructure as Code--Technology Review and
  Research Challenges}.
\newblock  (\bibinfo{year}{2025}).
\newblock


\bibitem[Parthasarathy et~al\mbox{.}(2025)]%
        {parthasarathy2025engineering}
\bibfield{author}{\bibinfo{person}{Kannan Parthasarathy},
  \bibinfo{person}{Karthik Vaidhyanathan}, \bibinfo{person}{Rudra Dhar},
  \bibinfo{person}{Venkat Krishnamachari}, \bibinfo{person}{Adyansh Kakran},
  \bibinfo{person}{Sreemaee Akshathala}, \bibinfo{person}{Shrikara Arun},
  \bibinfo{person}{Amey Karan}, \bibinfo{person}{Basil Muhammed},
  \bibinfo{person}{Sumant Dubey}, {et~al\mbox{.}}}
  \bibinfo{year}{2025}\natexlab{}.
\newblock \showarticletitle{Engineering LLM Powered Multi-Agent Framework for
  Autonomous CloudOps}. In \bibinfo{booktitle}{\emph{2025 IEEE/ACM 4th
  International Conference on AI Engineering--Software Engineering for AI
  (CAIN)}}. IEEE, \bibinfo{pages}{201--211}.
\newblock


\bibitem[Rahman et~al\mbox{.}(2018)]%
        {rahman2018bugs}
\bibfield{author}{\bibinfo{person}{Akond Rahman}, \bibinfo{person}{Sarah
  Elder}, \bibinfo{person}{Faysal~Hossain Shezan}, \bibinfo{person}{Vanessa
  Frost}, \bibinfo{person}{Jonathan Stallings}, {and} \bibinfo{person}{Laurie
  Williams}.} \bibinfo{year}{2018}\natexlab{}.
\newblock \showarticletitle{Bugs in infrastructure as code}.
\newblock \bibinfo{journal}{\emph{arXiv preprint arXiv:1809.07937}}
  (\bibinfo{year}{2018}).
\newblock


\bibitem[Rahman et~al\mbox{.}(2019)]%
        {rahman2019systematic}
\bibfield{author}{\bibinfo{person}{Akond Rahman}, \bibinfo{person}{Rezvan
  Mahdavi-Hezaveh}, {and} \bibinfo{person}{Laurie Williams}.}
  \bibinfo{year}{2019}\natexlab{}.
\newblock \showarticletitle{A systematic mapping study of infrastructure as
  code research}.
\newblock \bibinfo{journal}{\emph{Information and Software Technology}}
  \bibinfo{volume}{108} (\bibinfo{year}{2019}), \bibinfo{pages}{65--77}.
\newblock


\bibitem[{Red Hat}(2025)]%
        {ansible}
\bibfield{author}{\bibinfo{person}{{Red Hat}}.}
  \bibinfo{year}{2025}\natexlab{}.
\newblock \bibinfo{title}{Ansible}.
\newblock \bibinfo{howpublished}{\url{https://www.ansible.com/}}.
\newblock
\newblock
\shownote{Accessed: 2025-02-24}.


\bibitem[Roy et~al\mbox{.}(2024)]%
        {roy:2024:llm:rca}
\bibfield{author}{\bibinfo{person}{Devjeet Roy}, \bibinfo{person}{Xuchao
  Zhang}, \bibinfo{person}{Rashi Bhave}, \bibinfo{person}{Chetan Bansal},
  \bibinfo{person}{Pedro Las-Casas}, \bibinfo{person}{Rodrigo Fonseca}, {and}
  \bibinfo{person}{Saravan Rajmohan}.} \bibinfo{year}{2024}\natexlab{}.
\newblock \showarticletitle{Exploring LLM-Based Agents for Root Cause
  Analysis}. In \bibinfo{booktitle}{\emph{Companion Proceedings of the 32nd ACM
  International Conference on the Foundations of Software Engineering}} (Porto
  de Galinhas, Brazil) \emph{(\bibinfo{series}{FSE 2024})}.
\newblock
\href{https://doi.org/10.1145/3663529.3663841}{doi:\nolinkurl{10.1145/3663529.3663841}}


\bibitem[Schick et~al\mbox{.}(2023)]%
        {schick2023toolformer}
\bibfield{author}{\bibinfo{person}{Timo Schick}, \bibinfo{person}{Jane
  Dwivedi-Yu}, \bibinfo{person}{Roberto Dess{\`\i}}, \bibinfo{person}{Roberta
  Raileanu}, \bibinfo{person}{Maria Lomeli}, \bibinfo{person}{Eric Hambro},
  \bibinfo{person}{Luke Zettlemoyer}, \bibinfo{person}{Nicola Cancedda}, {and}
  \bibinfo{person}{Thomas Scialom}.} \bibinfo{year}{2023}\natexlab{}.
\newblock \showarticletitle{Toolformer: Language models can teach themselves to
  use tools}.
\newblock \bibinfo{journal}{\emph{Advances in Neural Information Processing
  Systems}}  \bibinfo{volume}{36} (\bibinfo{year}{2023}),
  \bibinfo{pages}{68539--68551}.
\newblock


\bibitem[Sheffler(2025)]%
        {sheffler2025approach}
\bibfield{author}{\bibinfo{person}{Thomas~J Sheffler}.}
  \bibinfo{year}{2025}\natexlab{}.
\newblock \showarticletitle{An Approach to Checking Correctness for Agentic
  Systems}.
\newblock \bibinfo{journal}{\emph{arXiv preprint arXiv:2509.20364}}
  (\bibinfo{year}{2025}).
\newblock


\bibitem[Shetty et~al\mbox{.}(2024)]%
        {shetty2024building}
\bibfield{author}{\bibinfo{person}{Manish Shetty}, \bibinfo{person}{Yinfang
  Chen}, \bibinfo{person}{Gagan Somashekar}, \bibinfo{person}{Minghua Ma},
  \bibinfo{person}{Yogesh Simmhan}, \bibinfo{person}{Xuchao Zhang},
  \bibinfo{person}{Jonathan Mace}, \bibinfo{person}{Dax Vandevoorde},
  \bibinfo{person}{Pedro Las-Casas}, \bibinfo{person}{Shachee~Mishra Gupta},
  {et~al\mbox{.}}} \bibinfo{year}{2024}\natexlab{}.
\newblock \showarticletitle{Building ai agents for autonomous clouds:
  Challenges and design principles}. In \bibinfo{booktitle}{\emph{Proceedings
  of the 2024 ACM Symposium on Cloud Computing}}. \bibinfo{pages}{99--110}.
\newblock


\bibitem[Su et~al\mbox{.}(2025)]%
        {su2025failure}
\bibfield{author}{\bibinfo{person}{Junhao Su}, \bibinfo{person}{Yuanliang Wan},
  \bibinfo{person}{Junwei Yang}, \bibinfo{person}{Hengyu Shi},
  \bibinfo{person}{Tianyang Han}, \bibinfo{person}{Junfeng Luo}, {and}
  \bibinfo{person}{Yurui Qiu}.} \bibinfo{year}{2025}\natexlab{}.
\newblock \showarticletitle{Failure Makes the Agent Stronger: Enhancing
  Accuracy through Structured Reflection for Reliable Tool Interactions}.
\newblock \bibinfo{journal}{\emph{arXiv preprint arXiv:2509.18847}}
  (\bibinfo{year}{2025}).
\newblock


\bibitem[Sun et~al\mbox{.}(2024)]%
        {sun-etal-2024-tools}
\bibfield{author}{\bibinfo{person}{Jimin Sun}, \bibinfo{person}{So~Yeon Min},
  \bibinfo{person}{Yingshan Chang}, {and} \bibinfo{person}{Yonatan Bisk}.}
  \bibinfo{year}{2024}\natexlab{}.
\newblock \showarticletitle{Tools Fail: Detecting Silent Errors in Faulty
  Tools}. In \bibinfo{booktitle}{\emph{Proceedings of the 2024 Conference on
  Empirical Methods in Natural Language Processing}},
  \bibfield{editor}{\bibinfo{person}{Yaser Al-Onaizan}, \bibinfo{person}{Mohit
  Bansal}, {and} \bibinfo{person}{Yun-Nung Chen}} (Eds.).
  \bibinfo{publisher}{Association for Computational Linguistics},
  \bibinfo{address}{Miami, Florida, USA}, \bibinfo{pages}{14272--14289}.
\newblock
\href{https://doi.org/10.18653/v1/2024.emnlp-main.790}{doi:\nolinkurl{10.18653/v1/2024.emnlp-main.790}}


\bibitem[Thiyagarajan et~al\mbox{.}(2024)]%
        {thiyagarajan2024ai}
\bibfield{author}{\bibinfo{person}{Gogulakrishnan Thiyagarajan},
  \bibinfo{person}{Vinay Bist}, {and} \bibinfo{person}{Prabhudarshi Nayak}.}
  \bibinfo{year}{2024}\natexlab{}.
\newblock \showarticletitle{AI-Driven Configuration Drift Detection in Cloud
  Environments}.
\newblock \bibinfo{journal}{\emph{Gogulakrishnan Thiyagarajan, Vinay Bist,
  Prabhudarshi Nayak.(2024). AI-Driven Configuration Drift Detection in Cloud
  Environments. International Journal of Communication Networks and Information
  Security (IJCNIS)}} \bibinfo{volume}{16}, \bibinfo{number}{5}
  (\bibinfo{year}{2024}), \bibinfo{pages}{721--743}.
\newblock


\bibitem[Verdet(2023)]%
        {verdet2023exploring}
\bibfield{author}{\bibinfo{person}{Alexandre Verdet}.}
  \bibinfo{year}{2023}\natexlab{}.
\newblock \bibinfo{booktitle}{\emph{Exploring security practices in
  infrastructure as code: An empirical study}}.
\newblock \bibinfo{publisher}{Ecole Polytechnique, Montreal (Canada)}.
\newblock


\bibitem[Vuddanti et~al\mbox{.}(2025)]%
        {vuddanti2025paladin}
\bibfield{author}{\bibinfo{person}{Sri~Vatsa Vuddanti}, \bibinfo{person}{Aarav
  Shah}, \bibinfo{person}{Satwik~Kumar Chittiprolu}, \bibinfo{person}{Tony
  Song}, \bibinfo{person}{Sunishchal Dev}, \bibinfo{person}{Kevin Zhu}, {and}
  \bibinfo{person}{Maheep Chaudhary}.} \bibinfo{year}{2025}\natexlab{}.
\newblock \showarticletitle{PALADIN: Self-Correcting Language Model Agents to
  Cure Tool-Failure Cases}.
\newblock \bibinfo{journal}{\emph{arXiv preprint arXiv:2509.25238}}
  (\bibinfo{year}{2025}).
\newblock


\bibitem[Wang(2022)]%
        {wang2022infrastructure}
\bibfield{author}{\bibinfo{person}{Rosemary Wang}.}
  \bibinfo{year}{2022}\natexlab{}.
\newblock \bibinfo{booktitle}{\emph{Infrastructure as Code, Patterns and
  Practices: With Examples in Python and Terraform}}.
\newblock \bibinfo{publisher}{Simon and Schuster}.
\newblock


\bibitem[Xiang et~al\mbox{.}(2025)]%
        {xiang2025automated}
\bibfield{author}{\bibinfo{person}{Yiming Xiang}, \bibinfo{person}{Zhenning
  Yang}, \bibinfo{person}{Jingjia Peng}, \bibinfo{person}{Hermann Bauer},
  \bibinfo{person}{Patrick Tser~Jern Kon}, \bibinfo{person}{Yiming Qiu}, {and}
  \bibinfo{person}{Ang Chen}.} \bibinfo{year}{2025}\natexlab{}.
\newblock \showarticletitle{Automated bug discovery in cloud
  infrastructure-as-code updates with llm agents}. In
  \bibinfo{booktitle}{\emph{2025 IEEE/ACM International Workshop on Cloud
  Intelligence \& AIOps (AIOps)}}. IEEE, \bibinfo{pages}{20--25}.
\newblock


\bibitem[Yang et~al\mbox{.}(2025a)]%
        {yang2025cloud}
\bibfield{author}{\bibinfo{person}{Zhenning Yang}, \bibinfo{person}{Archit
  Bhatnagar}, \bibinfo{person}{Yiming Qiu}, \bibinfo{person}{Tongyuan Miao},
  \bibinfo{person}{Patrick Tser Jern~Kon}, \bibinfo{person}{Yunming Xiao},
  \bibinfo{person}{Yibo Huang}, \bibinfo{person}{Martin Casado}, {and}
  \bibinfo{person}{Ang Chen}.} \bibinfo{year}{2025}\natexlab{a}.
\newblock \showarticletitle{Cloud infrastructure management in the age of ai
  agents}.
\newblock \bibinfo{journal}{\emph{ACM SIGOPS Operating Systems Review}}
  \bibinfo{volume}{59}, \bibinfo{number}{1} (\bibinfo{year}{2025}),
  \bibinfo{pages}{1--8}.
\newblock


\bibitem[Yang et~al\mbox{.}(2025b)]%
        {yang2025automated}
\bibfield{author}{\bibinfo{person}{Zhenning Yang}, \bibinfo{person}{Hui Guan},
  \bibinfo{person}{Victor Nicolet}, \bibinfo{person}{Brandon Paulsen},
  \bibinfo{person}{Joey Dodds}, \bibinfo{person}{Daniel Kroening}, {and}
  \bibinfo{person}{Ang Chen}.} \bibinfo{year}{2025}\natexlab{b}.
\newblock \showarticletitle{Automated Cloud Infrastructure-as-Code
  Reconciliation with AI Agents}.
\newblock \bibinfo{journal}{\emph{arXiv preprint arXiv:2510.20211}}
  (\bibinfo{year}{2025}).
\newblock


\bibitem[Yao et~al\mbox{.}(2023)]%
        {yao2023react}
\bibfield{author}{\bibinfo{person}{Shunyu Yao}, \bibinfo{person}{Jeffrey Zhao},
  \bibinfo{person}{Dian Yu}, \bibinfo{person}{Nan Du}, \bibinfo{person}{Izhak
  Shafran}, \bibinfo{person}{Karthik Narasimhan}, {and} \bibinfo{person}{Yuan
  Cao}.} \bibinfo{year}{2023}\natexlab{}.
\newblock \showarticletitle{React: Synergizing reasoning and acting in language
  models}. In \bibinfo{booktitle}{\emph{International Conference on Learning
  Representations (ICLR)}}.
\newblock
\showeprint[arxiv]{2210.03629}
\urldef\tempurl%
\url{https://openreview.net/pdf?id=WE_vluYUL-X}
\showURL{%
\tempurl}


\bibitem[Zhu et~al\mbox{.}(2025)]%
        {zhu2025llm}
\bibfield{author}{\bibinfo{person}{Kunlun Zhu}, \bibinfo{person}{Zijia Liu},
  \bibinfo{person}{Bingxuan Li}, \bibinfo{person}{Muxin Tian},
  \bibinfo{person}{Yingxuan Yang}, \bibinfo{person}{Jiaxun Zhang},
  \bibinfo{person}{Pengrui Han}, \bibinfo{person}{Qipeng Xie},
  \bibinfo{person}{Fuyang Cui}, \bibinfo{person}{Weijia Zhang},
  {et~al\mbox{.}}} \bibinfo{year}{2025}\natexlab{}.
\newblock \showarticletitle{Where LLM Agents Fail and How They can Learn From
  Failures}.
\newblock \bibinfo{journal}{\emph{arXiv preprint arXiv:2509.25370}}
  (\bibinfo{year}{2025}).
\newblock


\end{thebibliography}

\clearpage
\appendix

\section{Terraform Configuration Example}
\label{app:terraform}

This appendix provides a concrete \textsc{Terraform} configuration snippet illustrating how \ac{IaC} definitions can diverge from live infrastructure, along with representative examples of configuration drift arising during and after provisioning.

\begin{figure}[h]
\caption{A \textsc{Terraform} configuration snippet provisioning a web server and assigning it a security group.}
\label{lst:terraform}
\begin{lstlisting}[basicstyle=\small\ttfamily, numbers=left, breaklines=true]
resource "aws_instance" "web" {
  ami           = "ami-123"
  instance_type = "t2.micro"

  vpc_security_group_ids = [aws_security_group.web_sg.id]

  tags = {
    Name        = "web-server"
    Environment = "production"
  }
}
\end{lstlisting}
\end{figure}

The snippet in~\Cref{lst:terraform} declaratively defines a web server instance, including its operating system image, compute type, and network security group.
It also attaches identifying tags, which are commonly used to organize resources and express their intended role within the infrastructure.
This approach enables reproducibility, automation, version control, and consistency.

In practice, deployed resources frequently diverge from the IaC definition, which is known as \emph{configuration drift}~\cite{pahl2025infrastructure}.
Configuration drift can happen during provisioning, for instance when subtle bugs in IaC modules or provider APIs cause resources to be created with missing or inconsistent attributes.
For instance, an intermittent cloud API issue may result in the instance in~\Cref{lst:terraform} being provisioned without the intended security group due to a silent error in the backend, even though Terraform reports success.
Such mismatches are difficult to detect immediately because the \ac{IaC} tool may not report the underlying inconsistency.

Configuration drift can also occur after provisioning.
Manual hotfixes, automated scaling actions, system updates, or emergency interventions may modify configuration parameters outside the \ac{IaC} workflow.
An operator responding to an incident might manually update the instance's security group or temporarily open an SSH port.
Due to human error, these changes can remain in the live environment but are not captured in the \textsc{Terraform} file.
Over time, these discrepancies accumulate, making the operational state increasingly disconnected from its specification.

\section{Tool Reliability Illustrative Example}
\label{app:pingnode}

This appendix provides a worked example of how erroneous tool outputs can silently mislead an agent, motivating the reliability problem addressed by \sys.

\begin{figure}[h]
\caption{A simplified tool used by an agent to verify node reachability based on an expected IP mapping.}
\label{lst:pingnode}
\begin{lstlisting}[basicstyle=\small\ttfamily, numbers=left, breaklines=true, language=Python]
# Expected mapping between logical node identifiers and their IP addresses
nodes = {
    "0": "172.17.0.5",
    "1": "172.17.0.6",
}

# Tool invoked by the agent to check whether a node is reachable
def ping_node(node_id):
    return ping(nodes[node_id])
\end{lstlisting}
\end{figure}

\textbf{Illustrating example.}
Consider~\Cref{lst:pingnode} with a \texttt{ping\_node} function call that pings a node based on its identifier.
Each node identifier is mapped to an expected IP address and the agent verifies reachability of a node by invoking \texttt{ping\_node(id)}.
Under normal circumstances, this tool call returns correct information about the intended node.
However, drift or network reconfiguration may silently invalidate these assumptions.
For instance, the router might reassign IP addresses after a network event, or an operator may manually update network settings during an emergency, invalidating the IP-to-node mapping.
Because the tool returns syntactically correct output, the agent has no direct signal that anything is wrong.
For example, it concludes that node~1 is healthy even though the ping was forwarded to an unrelated device.

While the above example is straightforward, the underlying issue becomes far more severe in realistic cloud environments.
Modern agentic systems interact with dozens of heterogeneous tools, each with their own failure modes, silent inconsistencies, and assumptions.
As the number of tools and unknowns grows, the inability of the agent to distinguish genuine configuration drift from misleading but syntactically valid tool outputs increases.

\section{IaC Verification Workflow}
\label{app:workflow}

This appendix illustrates the end-to-end workflow of an \ac{LLM} agent verifying \ac{IaC}-defined infrastructure, as referenced in Section~\ref{sec:background}.

\begin{figure*}[h]
\centering
\includegraphics[width=\linewidth]{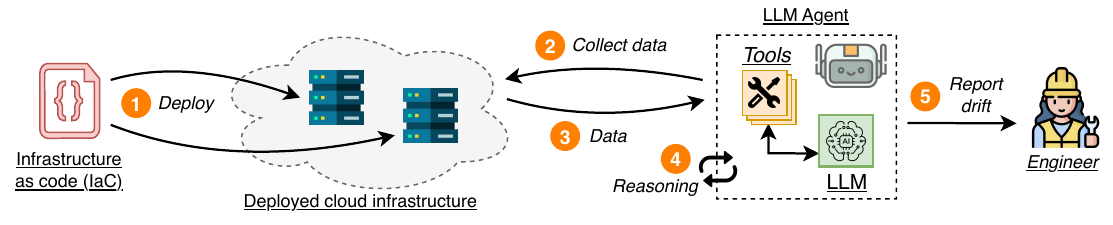}
\caption{The workflow of configuration drift detection by \ac{LLM} agents on infrastructure deployed by \ac{IaC}.}
\label{fig:motivation_workflow}
\end{figure*}

An engineer first deploys some \ac{IaC} specification (step~1).
Then, an \ac{LLM} agent collects data by using its available tools (steps~2 and~3), and reasons about the data (step~4) to identify potential issues.
This process may repeat for multiple steps.
Once the agent completes its task, it provides a report to the engineer who can further investigate identified issues.

\section{Detailed Efficiency Analysis}
\label{app:efficiency}

This appendix presents the full CDF distributions for step count and token usage of \sys and the ReAct baseline referenced in Section~\ref{ssec:efficiency}.

\begin{figure*}[h]
    \centering
    \includegraphics{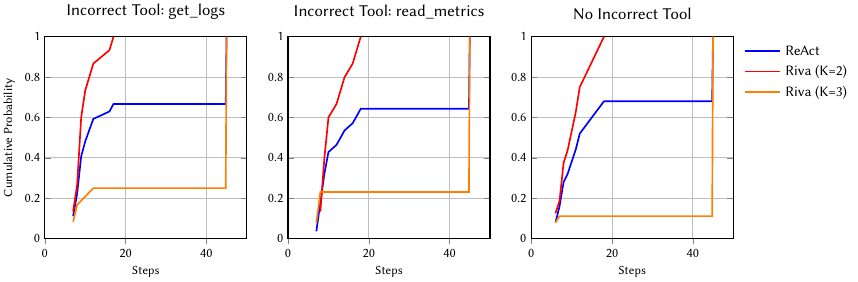}
    \caption{The distribution of number of steps required to complete tasks, for \sys and the ReAct agent, with and without erroneous tool responses.}
    \label{fig:cdf_steps}
\end{figure*}

\begin{figure*}[h]
    \centering
    \includegraphics{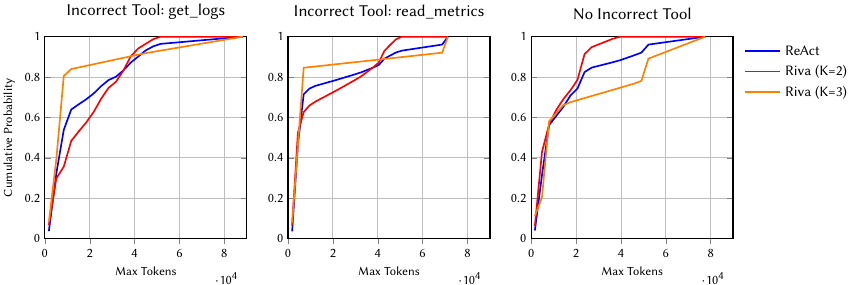}
    \caption{The distribution of maximum number of tokens required to complete tasks, for \sys and the ReAct agent, with and without erroneous tool responses.}
    \label{fig:cdf_tokens}
\end{figure*}

\Cref{fig:cdf_steps} (right) shows that when all tool responses are correct, 80\% of all tasks are completed in 15 steps with \sys (and $K=2$) while only 60\% of ReAct runs complete within this number of steps.
Moreover, the maximum number of steps with \sys ($K=2$) is 17 while 33\% of the runs with the ReAct agent require 45 steps, the maximum number of steps allowed before the task automatically aborts.
Similarly, \Cref{fig:cdf_tokens} (right) shows that the distribution of maximum number of tokens used per task for \sys ($K=2$) is within the range of the ReAct agent, although the maximum number of tokens for \sys ($K=2$) is \num{38000} tokens, compared to ReAct where a task requires at most \num{78000} tokens.
We attribute this to the two-agent design of \sys, which effectively distributes the information between the two agents, resulting in smaller context usage, reducing the maximum token count and lowering the chances of hallucination.
The smaller context sizes in \sys also reduce the number of incorrect tool calls per step, decreasing the amount of corrective steps the agent takes to fix tool calls.

In the presence of incorrect tools, \sys uses fewer steps than ReAct on average to finish a task (see~\Cref{fig:cdf_steps} left and middle).
In particular, \sys requires at most 17 steps with erroneous tools whereas 37\% of all ReAct tasks requires more than 17 steps.
However, \Cref{fig:cdf_tokens} (left and middle) shows that \sys uses a higher maximum number of tokens per task when tools are incorrect, though the tendency is inverted after 40000 tokens as the top 20\% of ReAct runs exceed the top 20\% of \sys runs.
Similarly to the experiment with only correct tools, the maximum value is much smaller than ReAct: \num{50000} tokens for \sys versus \num{90000} for ReAct.
Upon further inspection, \sys starts generating more commands when erroneous tools are present compared to when all tools provide correct outputs, resulting in higher token usage that allows the system to avoid the negative effects of incorrect tools.
ReAct, on the other hand, is not able to recover from incorrect results and exhausts all of its steps trying to reason over them.
Overall, the design of \sys allows for more effective use of its resources compared to the ReAct agent, particularly in the maximum number of tokens used to solve tasks.

\section{K Hyperparameter Sensitivity Analysis}
\label{app:k}

This appendix provides a detailed analysis of why \sys fails completely when $K=3$, as referenced in Section~\ref{ssec:k}.

\Cref{fig:cdf_steps} and \Cref{fig:cdf_tokens} show that $K=3$ leads to significantly different results: 83\% of all tasks use fewer than 15000 tokens while 77\% of all runs reach 45 steps, the maximum allowed in our evaluation.
Deeper analysis reveals that most executions fail at generating a third command.
In particular, AIOpsLab does not offer enough flexibility to the generator agent to produce more than 2 different tool calls for a specific goal.
As soon as the verifier agent tasks the generator agent to generate a third command, the generator continues trying until it reaches the maximum number of steps.
Moreover, as all generated tool calls are incorrect, they only produce a small number of tokens (an error message from the benchmark), explaining the small maximum token values observed.

\end{document}